\documentclass[12pt,preprint]{aastex}

\slugcomment{In Preparation for  ApJ, 2002}

\shorttitle{H$_2$O in Arcturus}

\shortauthors{ryde et al.}

\begin{document}


\title{Detection of Water Vapor in the Photosphere of
Arcturus}


\author{N. Ryde, D. L. Lambert, M. J. Richter\altaffilmark{1,2},
J. H. Lacy\altaffilmark{1}} \altaffiltext{1}{Visiting Astronomer
at the Infrared Telescope Facility, which is operated by the
University of Hawaii under contract from the National Aeronautics
and Space Administration.}\altaffiltext{2}{New affiliation:
Department of Physics, University of California, Davis, CA 95616}
\affil{McDonald Observatory and Department of Astronomy,
University of Texas at Austin, TX 78712-1083}
\email{ryde@astro.as.utexas.edu, dll@astro.as.utexas.edu,
richter@astro.as.utexas.edu, and lacy@astro.as.utexas.edu}







\begin{abstract}
We report detections of pure rotation lines of OH and H$_2$O in
the K1.5 III red-giant star Arcturus ($\alpha$ Bo\"otis) using
high-resolution, infrared spectra covering the regions
$806-822\,\mbox{cm$^{-1}$}$ ($12.2-12.4\,\mbox{$\mu$m}$) and
$884-923\,\mbox{cm$^{-1}$}$ ($10.8-11.3\,\mbox{$\mu$m}$). Arcturus
is the hottest star yet to show water-vapor features in its
disk-averaged spectrum. We argue that the water vapor lines
originate from the photosphere, albeit in the outer layers.
We are able to predict the observed strengths of OH
 and H$_2$O lines satisfactorily after
lowering the temperature structure of the very outer parts of the
photosphere ($\log \tau_{500}=-3.8$ and beyond) compared to a
flux-constant, hydrostatic, standard {\sc marcs} model
photosphere. Our new model is consistently calculated including
chemical equilibrium and radiative transfer from the given
temperature structure. Possible reasons for a temperature decrease
in the outer-most parts of the photosphere and the assumed
break-down of the assumptions made in classical model-atmosphere
codes are discussed.

\end{abstract}



\section{INTRODUCTION}

Water is one of the most abundant molecules in the atmospheres of
oxygen-rich, late-type stars. It is a dominant source of opacity
in the infrared and, therefore, plays an important role in
determining the structure of cool photospheres. Water is prominent
the spectra of brown dwarfs, M-dwarfs and Mira variable stars.
Water vapor is, however, not expected in the photospheres of giant
stars hotter than late M-type [see, for example, the discussion in
\citet{tsuji_2001}]. Here, we will present evidence for water
vapor in the mid-infrared, photospheric spectrum of the K1.5
III\footnote{http://simbad.u-strasbg.fr/} red giant Arcturus
($\alpha$ Bo\"{o}tis), the hottest star yet to have shown
water-vapor absorption lines.

Mid-infrared, water-vapor lines have earlier been observed in
sunspot spectra. \citet{wallace_science} reported on the
identification of numerous, resolved water-vapor lines in a
sunspot spectrum of the N-band ($760-1233\,\mbox{cm$^{-1}$}$).
\citet{poly_3} were later able to assign quantum numbers to these
lines, which is a very difficult task due to the great complexity
of the water vapor spectrum.
Wallace et al. \nocite{wallace_science} derived an effective
temperature of the observed sunspot of approximately
$3300\,\mbox{K}$, which corresponds to a spectral class of M2-5.
\citet{antares} identified water-vapor lines at
$811-819\,\mbox{cm$^{-1}$}$ of the  M-type supergiants $\alpha$
Orionis (Betelgeuse; M1$^2$) and $\alpha$ Scorpii (Antares;
M1.5$^2$). These were observed at a resolution of $R\approx
10,000$. Jennings \& Sada \nocite{antares} modelled their
observations with a plane-parallel, isothermal, single layer close
to the location of the onset of the chromospheric temperature
rise.

In order to explain several discrepancies between models and
near-infrared observations of late K and M giants and supergiants,
also showing water vapor, Tsuji and collaborators \citep[see for
example][]{tsuji_1997,tsuji_1998,tsuji_ny,tsuji_2000,yam_99} have
introduced the idea of a stationary, warm envelope situated at a
distance of a few stellar radii above the photosphere but interior
to the cool, expanding circumstellar-shell. This previously
undetected envelope (called the MOLsphere) is considered to
contain water vapor at temperatures of $1000$--$2000\,\mbox{K}$
\citep{tsuji_1997}, resulting in non-photospheric signatures in IR
spectra of M giants. The envelope was neither theoretically
predicted nor has it as yet received a theoretical explanation,
but the water lines and bands seem to be a common feature of M
supergiants and M giants
\citep{tsuji_1998,matsuura}. Numerous pieces of evidence have been
presented in favor of this idea
\citep[see for example][]{tsuji_1997, tsuji_1998,tsuji_2000}.
Here, we will, however, argue that the water-vapor lines detected
in the observations of $\alpha$ Boo
are not from a MOLsphere but of photospheric origin, a possibility
anticipated by \citet{tsuji:88}.

Very little work has been done on resolved molecular lines in the
mid-infrared region. Our analysis is based on spectra of Arcturus
obtained with the
{\sc texes} spectrograph, a unique, high-resolution, mid-infrared
spectrograph.  This is the first time, to our knowledge, this
wavelength region has been observed at  high-resolution with high
sensitivity. In Section 2 we present the observations and in Sect.
3 we identify the observed features and present our line lists
which are used in the generation of synthetic spectra, a process
described in Sect. 4. Section 5 we discuss our findings.

\section{OBSERVATIONS}

\begin{figure}
\caption{Part of the observed Arcturus spectrum
at $815-817\,\mbox{cm$^{-1}$}$ showing two strong
OH($v=0\leftarrow0$) lines and several H$_2$O lines. Data from
four orders are shown, with three inter-order gaps visible.
\label{figsample}}
\end{figure}


\def\cm1{\,\mbox{cm$^{-1}$}}
\def\kms{\,\mbox{km\,s$^{-1}$}}

We observed Arcturus at $806.3-821.4 \,\mbox{cm$^{-1}$}$,
$883.8-901.6 \,\mbox{cm$^{-1}$}$, and $903.6-922.7
\,\mbox{cm$^{-1}$}$
on 2001 February 2, 4, and 3, respectively,
with the Texas Echelon-Cross-Echelle Spectrograph ({\sc texes};
Lacy et al. 2002)\nocite{texes} mounted on the 3 meter NASA
Infrared Telescope Facility ({\sc irtf}). {\sc texes} is a
ground-based prototype of {\sc exes}, a mid-infrared
($350-1800\,\mbox{cm$^{-1}$}$) spectrograph designed for use on
{\sc sofia} (the Stratospheric Observatory For Infrared
Astronomy). We observed with a resolving power of
$\tilde{\nu}/\Delta\tilde{\nu}\approx 80,000$, $\tilde{\nu}$ being
the wavenumber.  A sample of the observed spectrum at
$815-817\,\mbox{cm$^{-1}$}$ is shown in Figure \ref{figsample}.
The lines identified by comparing with the sunspot spectrum
\citep{solspectrum} are pure rotation lines of water vapor and OH.
The rotational lines of OH come in quartets.
We have identified 17 water-vapor lines in the observed spectra.
Several aluminum, silicon, and magnesium lines are also detected,
most of which are seen in emission. These metal lines
will be dealt with in a subsequent paper.

{\sc texes} has a 0.9 meter long, aluminum echelon grating (blaze
angle $84^\circ$ or R10) and a choice of an echelle or first order
grating for cross-dispersion.  The detector array is a 256$^2$
pixel, Si:As IBC from the Raytheon Infrared Center for Excellence
and designed for low backgrounds such as in the Space Infrared
Telescope Facility ({\sc sirtf}) or high resolution spectroscopy.
We used the {\sc texes} `hi-lo' spectral mode with the echelon
grating cross-dispersed by the first order grating. This mode
provides large spectral coverage at the cost of a short slit. It
is appropriate for observing bright point sources in stable
atmospheric conditions, as were present during these observations.
The on-source integration times ranged from 1-2 minutes. Arcturus
and Sirius, a hot star intended to be an atmospheric calibrator,
were observed in the same fashion: four to six (depending on the
night) 1 second integrations on the target followed by the same
number of integrations on blank sky five arcseconds away.  At this
distance from bright stars there is an insignificant amount of
scattered light. We repeated this cycle either 16 or 32 times to
reach the final integration time.  As part of each observation
set, we also observed a telescope-temperature black-body  and a
sky frame for flat-fielding and wavelength calibration, as
described below.

The data were reduced using the standard {\sc texes} pipeline with
flat-fielding, atmospheric correction, wavelength calibration
using atmospheric lines, and optimal extraction of the spectrum.
We create a flat field and first order atmospheric calibration
frame by subtracting the sky frame from the black-body frame. The
sky frame contains emission from telluric atmospheric lines
according to the line optical depth and species temperature
structure within our atmosphere.  This provides a measure of the
instrumental response to uniform illumination where the atmosphere
is clean and gives an approximate correction for regions affected
by the telluric atmosphere. We establish a frequency scale by
specifying the pixel value and frequency of one atmospheric line
in one order and calculating, based on our understanding of the
{\sc texes} optical components, the frequency scale for each order
on the array.  This process is subject to a systematic offset of
as much as a few $\kms$ if the pixel value is in error such as
might happen when the only telluric lines available are fairly
broad with poorly defined minima or if the only lines fall near
ends of echelon orders. Normally, though, the accuracy is
estimated to be on the order of $\pm0.5~\kms$, since one pixel
represents roughly $0.9~\kms$.

After reading the data into software, we remove background
emission by taking the difference of target frames and blank sky
frames. After flat-fielding, we re-sample the array so that the
spectral and spatial dimensions are orthogonal and run along rows
and columns. We use the spatial profile of the star to construct a
weighting function for optimal extraction of the stellar flux.  We
then have a 1D spectrum of the star.

Unfortunately, we have found that the black-body frame can result
in a non-linear array response causing structure across the orders
and improper atmospheric correction. We normalized each order by
fitting fourth-order polynomials to the continuum and dividing. In
a few orders, for example near $808\cm1$ and $814\cm1$, the
normalization procedure is hampered by stellar and telluric lines.
Because the signal-to-noise (S/N) in Arcturus was so high,
approximately four times the S/N in the atmospheric calibrator,
Sirius, we chose to make final atmospheric corrections using an
atmospheric modelling program written by Erik Grossman (private
communication). After a general correction, a second correction
was made for just the telluric water lines since water is highly
variable in Earth's atmosphere. Since this spectral region is
relatively free from telluric lines, the atmospheric correction
should not significantly alter our results.

The array also seems to exhibit charge trapping after bright
illuminations.  The trapped charges are released during subsequent
frames with a several second time constant, resulting in slightly
elevated background levels.  We simply ignore the first target-sky
pair of each observation cycle after observing the black-body
calibrator, where this effect is most prominent.

\section{FEATURE IDENTIFICATION AND LINE-LISTS}

Arcturus has a well determined and stable radial velocity which we
exploit to check the attribution of the pure rotation OH and
H$_2$O lines to the photosphere. The heliocentric velocity is
$-5.2 \,\mbox{km\,s$^{-1}$}$ \citep{evans} stable to about
$0.1\,\mbox{km\,s$^{-1}$}$ \citep{hc:93}. Our measured line
positions are corrected for this velocity and the contribution
from the Earth's velocity in order to obtain velocity shifts with
respect to the photospheric velocity.


\subsection{Vibration-Rotation OH Lines}

The OH radical is an expected photospheric constituent.
Vibration-rotation lines from the molecule's ground state are
securely identified in the infrared Arcturus Atlas (Hinkle et al.
1995a,b). We use a selection of the first-overtone lines near
$1.6\,\mbox{$\mu$m}$ to establish the photospheric contribution to
the pure rotation OH lines. In this section, we compare
vibration-rotation and pure rotation lines with respect to
velocity and line width. We later discuss the observed line
strengths.

\begin{figure}
\rotate \caption{Part of the analyzed Arcturus
spectrum around 1.6\mbox{$\mu$m}. The OH($v'-v''=2-0$) lines
are fitted well with a standard {\sc marcs} photospheric model.
The data are plotted with crosses and the model is shown with a
full line. Several metal lines are present. No attempt has,
however, been made to fit the abundances or uncertain $gf$-values
of individual metal lines, nor has an attempt been made to make a
complete line list for the region. The focus has been on the
vibration-rotation OH lines. The regularly spaced narrow spikes in
the spectrum are residuals from telluric water vapor lines after
division of a telluric spectrum. \label{fig4}}
\end{figure}

Figure \ref{fig4} shows several OH lines from the Hinkle et al.
Atlas. Laboratory line positions from Goldman et al. (1998)
\nocite{gold} via M\'{e}len et al. (1995)\nocite{melen} are
accurate to about $0.0038\,\mbox{cm$^{-1}$}$ or
$0.2\,\mbox{km\,s$^{-1}$}$. The data in  Table \ref{OH_1} are
  measured off the Arcturus
  Atlas. 
Also presented are line data from the \citet{gold} line-list.
Oscillator strengths given in the Table as $\log gf$-values are
taken from Goldman et al. (1998) and are considered accurate to
about  $10 \%$. Observed line positions scatter about the adopted
photospheric velocity: the mean velocity shift is $\langle\Delta
\tilde{\nu}_\mathrm{OH(6400)}\rangle = 0.1 \pm 0.2
\,\mbox{km\,s$^{-1}$}$. The mean measured width (FWHM) is $8.8 \pm
0.3\,\mbox{km\,s$^{-1}$}$ (see Table \ref{OH_1}).
The Atlas resolution corresponds to a line width of about
$1.9\,\mbox{km\,s$^{-1}$}$ implying that the true width of the OH
lines is approximately $\sqrt{8.8^2- 1.9^2} =
8.6\pm0.4\,\mbox{km\,s$^{-1}$}$.



\begin{table*}
  \caption{Observational data of OH,  vibration-rotational  lines at $6350-6430\,\mbox{cm$^{-1}$}$.
  }
  \label{obs_2}
  \label{OH_1}
  \begin{center}
  \begin{tabular}{llllllll} \hline
  \noalign{\smallskip}
  \multicolumn{1}{c}{$\tilde{\nu}_\mathrm{lab}$}     &
  \multicolumn{1}{c}{$\Delta\tilde{\nu}$} &
  \multicolumn{1}{c}{equivalent} &
  \multicolumn{1}{c}{FWHM} &
    \multicolumn{1}{c}{$E''_\mathrm{exc}$} &
  \multicolumn{1}{c}{$\log gf$} &
  \multicolumn{1}{c}{$v'-v''$}    &
  \multicolumn{1}{c}{Lower level
}

  \\
   \multicolumn{1}{c}{ } &
   \multicolumn{1}{c}{ } &
   \multicolumn{1}{c}{width} &
   \multicolumn{1}{c}{ } &
   \multicolumn{1}{c}{ } &
   \multicolumn{1}{c}{ } &
   \multicolumn{1}{c}{ } &
   \multicolumn{1}{c}{ }
 \\
   \multicolumn{1}{c}{[cm$^{-1}$]} &
   \multicolumn{1}{c}{[km\,s$^{-1}$]} &
   \multicolumn{1}{c}{[10$^{-3}$\,cm$^{-1}$]} &
   \multicolumn{1}{c}{[km\,s$^{-1}$] } &
      \multicolumn{1}{c}{[eV] } &
   \multicolumn{1}{c}{ } &
   \multicolumn{1}{c}{ } &
   \multicolumn{1}{c}{ }

  \\

 \noalign{\smallskip}
  \hline
  \noalign{\smallskip}
6355.363 
&  0.05   &  55.0 &  8.0  &  0.354 & -5.24&  2-0 & P$_\mathrm{1f}12.5$ \\
6356.868 
&  1.2    &  72.3 &  10.7 &  0.353 & -5.24&  2-0 & P$_\mathrm{1e}12.5$   \\
6359.707 
&  -0.15  &  45.4 &  7.6  &  0.358 & -5.27&  2-0 & P$_\mathrm{2e}11.5$ \\
6360.749 
&  -1.2   &  63.7 &  9.4  &  0.357 & -5.27&  2-0 & P$_\mathrm{2f}11.5$ \\
6386.612 
&  -0.32  &  46.8 &  10.4  &  0.534 & -5.17 & 3-1 & P$_\mathrm{1f}6.5$  \\
6387.258 
&  -0.11  &  40.4 & 7.8  &  0.534 & -5.17 & 3-1 & P$_\mathrm{1e}6.5$ \\
6397.266 
&   0.46  &  47.1 & 9.3  &  0.541 & -5.24 & 3-1 & P$_\mathrm{2e}5.5$ \\
6397.555 
&   0.49  & 42.1 &  8.1  &  0.541 & -5.24 & 3-1 & P$_\mathrm{2f}5.5$ \\
6419.992 
&   0.70  &  57.7 & 8.9  &  0.300 & -5.29 & 2-0 & P$_\mathrm{1f}11.5$ \\
6421.354 
&   0.13  &  51.9 & 8.1  &  0.299 & -5.29 & 2-0 & P$_\mathrm{1e}11.5$  \\
6424.877 
&   0.00  &  53.7 & 8.0   &  0.304 & -5.33 & 2-0 & P$_\mathrm{2e}10.5$\\
  \noalign{\smallskip}
  \hline
  \end{tabular}
  \end{center}
  \smallskip
\end{table*}

\begin{table*}
  \caption{Observational data and the line list of the water-vapor lines for the $806-822\,\mbox{cm$^{-1}$}$
  region  from 2001 February 2.
}
  \label{water_12um}
  \begin{center}
  \begin{tabular}{llllllll} \hline
  \noalign{\smallskip}
 \multicolumn{1}{c}{$\tilde{\nu}_\mathrm{lab}$}     &
  \multicolumn{1}{c}{$\Delta\tilde{\nu}$\tablenotemark{a}\tablenotetext{a}{A colon (:) marks measured values with large
uncertainties.}} &
  \multicolumn{1}{c}{equivalent} &
  \multicolumn{1}{c}{FWHM\tablenotemark{a}} &
   \multicolumn{1}{c}{$E''_\mathrm{exc}$} &
  \multicolumn{1}{c}{$\log gf$} &
  \multicolumn{1}{c}{$J'$$K'_a$$K'_c$$J''$$K''_a$$K''_c$}    &
  \multicolumn{1}{c}{$v_1v_2v_3$}
  \\
   \multicolumn{1}{c}{ } &
 \multicolumn{1}{c}{ } &
  \multicolumn{1}{c}{width\tablenotemark{a}} &
   \multicolumn{1}{c}{ } &
   \multicolumn{1}{c}{ } &
   \multicolumn{1}{c}{ } &
   \multicolumn{1}{c}{ } &
   \multicolumn{1}{c}{ }
    \\
   \multicolumn{1}{c}{[cm$^{-1}$]} &
   \multicolumn{1}{c}{[km\,s$^{-1}$]} &
   \multicolumn{1}{c}{[$10^{-3}$\,cm$^{-1}$]} &
   \multicolumn{1}{c}{[km\,s$^{-1}$] } &
   \multicolumn{1}{c}{[eV] } &
   \multicolumn{1}{c}{ } &
   \multicolumn{1}{c}{ } &
   \multicolumn{1}{c}{ }
  \\
 \noalign{\smallskip}
  \hline
 \noalign{\smallskip}
813.75067   
& 0.18: & 0.79:
&  5.7:
& 0.833 & -1.90  & 23     8   15   22  7    16 & (000) \\
815.30059   
& 0.54 & 1.3
&  7.1
& 0.498 & -2.51  & 18    7   12   17  4   13 & (000) \\
815.897\tablenotemark{b}\tablenotetext{b}{Uncertain assignment of the quantum numbers for the states of the transition.}  & & &  
& 1.396 & -1.00  & 21    21   0   20   20   1 & (010) \\
815.900\tablenotemark{b}   
& -0.74 & 0.85
&  7.8
& 1.396 & -1.48  & 21    21    0   20     20   1 & (010) \\
816.45026   
& 0.25: & 0.26:
&  7.2:
& 0.398 & -3.21  & 17   5    13   16  2    14 & (000) \\
816.68703   
& 0.34 &1.8
&  9.3 
& 1.014 & -1.35  & 24    12   13  23  11   12 & (000) \\
817.138   &  & & 
& 1.206 & -1.66  &     &  \\
817.15695  
& 0.72 & 1.6
&  9.1
& 1.206 & -1.18  & 21    16   5    20  15   6  & (010) \\
817.20881 
& -0.69: & 1.1:
&  10.5: 
& 1.014 & -1.83  & 24    12 12   23  11   13 & (000) \\
817.209   & &  & 
& 1.460 & -1.55  &       &  \\
818.4238\tablenotemark{c}\tablenotetext{c}{Assignments from Tsuji (2000)}  & &  & 
& 1.029 & -1.66  & 22    16  6    21  15    7 & (000) \\
818.4247\tablenotemark{c}   
& 0.022 & 2.6
&  8.0 
& 1.029 & -1.19  & 22    16   7    21  15    6 & (000) \\
819.93233  
& 0.089 & 0.78
&  6.6 
& 1.050 & -1.42  & 25    11   14   24 10  15 & (000) \\
\noalign{\smallskip}
  \hline
  \end{tabular}
  \end{center}
  \smallskip

\end{table*}

\begin{table*}
  \caption{Observational data and the line list of the water-vapor lines for the $884-925\,\mbox{cm$^{-1}$}$ region
   from 2001 February 3 and 4.
  }
  \label{water_11um}
  \begin{center}
  \begin{tabular}{llllllll} \hline
  \noalign{\smallskip}
 \multicolumn{1}{c}{$\tilde{\nu}_\mathrm{lab}$}     &
  \multicolumn{1}{c}{$\Delta\tilde{\nu}$\tablenotemark{a}} &
  \multicolumn{1}{c}{Equivalent} &
  \multicolumn{1}{c}{FWHM\tablenotemark{a}} &
   \multicolumn{1}{c}{$E''_\mathrm{exc}$} &
  \multicolumn{1}{c}{$\log gf$} &
  \multicolumn{1}{c}{$J'$$K'_a$$K'_c$$J''$$K''_a$$K''_c$}    &
  \multicolumn{1}{c}{$v_1v_2v_3$}
  \\
   \multicolumn{1}{c}{} &
   \multicolumn{1}{c}{} &
   \multicolumn{1}{c}{width\tablenotemark{a}} &
   \multicolumn{1}{c}{ } &
   \multicolumn{1}{c}{ } &
   \multicolumn{1}{c}{ } &
   \multicolumn{1}{c}{ } &
   \multicolumn{1}{c}{ }
 \\
   \multicolumn{1}{c}{[cm$^{-1}$]} &
   \multicolumn{1}{c}{[km\,s$^{-1}$]} &
   \multicolumn{1}{c}{[$10^{-3}$\,cm$^{-1}$]} &
   \multicolumn{1}{c}{[km\,s$^{-1}$] } &
   \multicolumn{1}{c}{[eV] } &
   \multicolumn{1}{c}{ } &
   \multicolumn{1}{c}{ } &
   \multicolumn{1}{c}{ }
  \\
 \noalign{\smallskip}
  \hline
 \noalign{\smallskip}
886.042   & & &
& 1.265 & -1.60  &      &  \\
886.04332 
& 1.61 & 1.0
&  8.4
& 1.265 & -1.12  &  25   17    8    24    16     9 & (000) \\
887.40891  
& 1.34 & 1.1
&  8.4
& 0.602 & -2.40  &  20   7  14   19     4    15 & (000) \\
894.63740  
& 1.96 & 0.94
&  7.9 
& 0.626 & -2.47  &  20    8   13    19    5    14 & (000) \\
894.647    & & &  
& 1.950 & -0.97  &   & \\
  \hline
 \noalign{\smallskip}
904.62103     
& -2.06: & 0.97:
&  8.3: 
& 0.804 & -2.16  &  23    7   16   22    6   17 & (000) \\
904.625   & & &     
& 1.505 & -1.11  &     &  \\
904.67827   & : &   : & :   
& 0.571 & -2.98  &  20    5   15    19    4    16 & (000) \\
 910.71027  
& 1.63: & 1.1:
&  8.8: 
 & 0.408 & -2.99  & 18     4    15   17     1   16 & (000) \\
 911.23428   
& 0.040 & 0.78
&  7.3 
 & 0.489 & -2.71  & 19     4    15  18     3   16 & (000) \\
 914.60786    & : & : & : 
 & 0.489 & -3.18  & 19     5    15   18     2   16 & (000) \\
\noalign{\smallskip}
  \hline
  \end{tabular}
  \end{center}
  \smallskip
  \tablenotetext{a}{A colon (:) marks measured values with large
uncertainties.}
\end{table*}

\begin{table*}
  \caption{Observational data of pure rotational, OH lines at
  $884-925\,\mbox{cm$^{-1}$}$ from 2001 February 3 and 4.
%
}
  \label{OH_2}
  \begin{center}
  \begin{tabular}{llllllll} \hline
  \noalign{\smallskip}
 \multicolumn{1}{c}{$\tilde{\nu}_\mathrm{lab}$}     &
  \multicolumn{1}{c}{$\Delta\tilde{\nu}$\tablenotemark{a}\tablenotetext{a}{A colon (:) marks measured values with large
uncertainties.}} &
  \multicolumn{1}{c}{Equivalent} &
  \multicolumn{1}{c}{FWHM\tablenotemark{a}} &
  \multicolumn{1}{c}{$E''_\mathrm{exc}$} &
  \multicolumn{1}{c}{$\log gf$} &
  \multicolumn{1}{c}{$v'-v''$}    &
  \multicolumn{1}{c}{Lower level}
  \\
   \multicolumn{1}{c}{ } &
   \multicolumn{1}{c}{ } &
   \multicolumn{1}{c}{width\tablenotemark{a}} &
   \multicolumn{1}{c}{ } &
   \multicolumn{1}{c}{ } &
   \multicolumn{1}{c}{ } &
   \multicolumn{1}{c}{ } &
   \multicolumn{1}{c}{ }
  \\
   \multicolumn{1}{c}{[cm$^{-1}$]} &
   \multicolumn{1}{c}{[km\,s$^{-1}$]} &
   \multicolumn{1}{c}{[$10^{-3}$\,cm$^{-1}$]} &
   \multicolumn{1}{c}{[km\,s$^{-1}$] }&
   \multicolumn{1}{c}{[eV] } &
   \multicolumn{1}{c}{ } &
   \multicolumn{1}{c}{ } &
   \multicolumn{1}{c}{ }
  \\
 \noalign{\smallskip}
  \hline
 \noalign{\smallskip}
883.9131 
&   1.1  & 9.5 & 10.3 & 1.614 & -1.50 & 0-0 & R$_\mathrm{2e}26.5$ \\ 
884.6252 
&   1.7  & 4.3 &   8.8 & 2.200 & -1.45 & 1-1 & R$_\mathrm{1e}29.5$ \\
885.2840 
&    1.7 & 3.6 &   8.5 & 2.205 & -1.46 & 1-1 & R$_\mathrm{2e}28.5$ \\
885.6448 
&    1.1 & 4.8 &   9.9 & 2.203 & -1.45 & 1-1 & R$_\mathrm{1f}29.5$ \\
891.4516  & 2.0: & 1.3:  & :
& 2.872 & -1.43 & 2-2 & R$_\mathrm{2f}31.5$ \\
892.1655  & 1.0: & 0.98:  & :
& 2.874 & -1.43 & 2-2 & R$_\mathrm{2e}31.5$ \\
892.4433   & 1.4: & 1.3: & 7.8:
& 2.873 & -1.41 & 2-2 & R$_\mathrm{1f}32.5$ \\
 \noalign{\smallskip}
  \hline
 \noalign{\smallskip}
903.7739 
&    -1.3 & 31.6 & 10.5 & 1.721 & -1.47 & 0-0 & R$_\mathrm{2f}27.5$ \\ 
904.0414 
&    -0.16 & 31.9 & 10.6 & 1.718 & -1.45 & 0-0 & R$_\mathrm{1e}28.5$ \\ 
904.8001 
&    0.24 & 30.3 & 10.0 & 1.723 & -1.47 & 0-0 & R$_\mathrm{2e}27.5$ \\ 
904.8011  & \multicolumn{2}{l}{blended with OH(0-0)} & 
& 2.982 & -1.41 & 2-2 & R$_\mathrm{2f}32.5$ \\
905.0040  & -0.42: & 1.4: &: 
& 2.980 & -1.39 & 2-2 & R$_\mathrm{1e}33.5$ \\
905.1867 
&   -0.97  & 30.8 & 10.2 & 1.722 & -1.45 & 0-0 & R$_\mathrm{1f}28.5$ \\ 
905.4647  & -2.0: & 0.69: &: 
& 2.985 & -1.41 & 2-2 & R$_\mathrm{2e}32.5$ \\
905.7167  & -0.64:  & 1.9: & :
& 2.983 & -1.39 & 2-2 & R$_\mathrm{1f}33.5$ \\
917.0493 & -2.2: & 1.0: & :& 3.096&-1.38  & 2-2 &  R$_\mathrm{2f}33.5$\\
917.2380 & 0.53: & 0.96: & :&3.097 & -1.39 & 2-2 & R$_\mathrm{1e}34.5$ \\
917.6575 & -1.6: & 0.91: & :&3.092 & -1.38 & 2-2 & R$_\mathrm{2e}33.5$ \\
917.8851 & 0.81: & 1.4: & :&3.095 & -1.39 & 2-2 & R$_\mathrm{1f}34.5$ \\
918.8106 
&   -1.0: & 3.2 &   8.0: & 2.424 & -1.42 & 1-1 & R$_\mathrm{2f}30.5$  \\
919.0291 
&   -1.2  &  3.1 & 8.1 & 2.421 & -1.40 & 1-1 & R$_\mathrm{1e}31.5$  \\
919.6607 
&   -0.76: & 4.4 &  10.7: & 2.426 & -1.42 & 1-1 & R$_\mathrm{2e}30.5$  \\
919.9632 
&   0.046 & 3.3 &  8.8 & 2.425 & -1.40 & 1-1 & R$_\mathrm{1f}31.5$  \\
\noalign{\smallskip}
  \hline
  \end{tabular}
  \end{center}
  \smallskip
\end{table*}

\begin{table*}
  \caption{
Observational data of pure rotational, OH lines at
  $806-822\,\mbox{cm$^{-1}$}$ from 2001 February 2.
  }
  \label{OH_3}
  \label{obs_3}
  \begin{center}
  \begin{tabular}{llllllll} \hline
  \noalign{\smallskip}
 \multicolumn{1}{c}{$\tilde{\nu}_\mathrm{lab}$}     &
  \multicolumn{1}{c}{$\Delta\tilde{\nu}$\tablenotemark{a}\tablenotetext{a}{A colon (:) marks measured values with large
uncertainties.}} &
  \multicolumn{1}{c}{Equivalent} &
  \multicolumn{1}{c}{FWHM\tablenotemark{a}} &
  \multicolumn{1}{c}{$E''_\mathrm{exc}$} &
  \multicolumn{1}{c}{$\log gf$} &
  \multicolumn{1}{c}{$v'-v''$}    &
  \multicolumn{1}{c}{Lower level}
  \\
   \multicolumn{1}{c}{ } &
   \multicolumn{1}{c}{ } &
   \multicolumn{1}{c}{width\tablenotemark{a}} &
   \multicolumn{1}{c}{ } &
   \multicolumn{1}{c}{ } &
   \multicolumn{1}{c}{ } &
   \multicolumn{1}{c}{ } &
   \multicolumn{1}{c}{ }
  \\
   \multicolumn{1}{c}{[cm$^{-1}$]} &
   \multicolumn{1}{c}{[km\,s$^{-1}$]} &
   \multicolumn{1}{c}{[$10^{-3}$\,cm$^{-1}$]} &
   \multicolumn{1}{c}{[km\,s$^{-1}$] }&
   \multicolumn{1}{c}{[eV] } &
   \multicolumn{1}{c}{ } &
   \multicolumn{1}{c}{ } &
   \multicolumn{1}{c}{ }
  \\
 \noalign{\smallskip}
  \hline
 \noalign{\smallskip}
808.9456 
&  0.03:  &  2.5: &  8.1: & 2.349 & -1.54 & 2-2 & R$_\mathrm{2f}26.5$ \\
809.2817 
&  -0.75  &  2.7 & 8.3 & 2.346 & -1.52 & 2-2 & R$_\mathrm{1e}27.5$ \\
809.8358 
&  -0.07 &  2.8 & 9.1 & 2.351 & -1.54 & 2-2 & R$_\mathrm{2e}26.5$ \\
810.2738 
&  -0.45:  &  2.6: & 8.0: & 2.349 & -1.52 & 2-2 & R$_\mathrm{1f}27.5$ \\
814.3245 
&  -0.50 & 11.5 & 11.1 & 1.300 & -1.59 & 0-0 & R$_\mathrm{2f}23.5$ \\ 
814.7280 
&   -1.06:  & 12.1 & 11.7 & 1.297 & -1.57 & 0-0 & R$_\mathrm{1e}24.5$ \\ 
815.4032 
&   -0.79: & 10.4 & 10.7 & 1.302 & -1.58 & 0-0 & R$_\mathrm{2e}23.5$ \\ 
815.9535 
&   -0.92 & 11.7 & 11.4 & 1.300 & -1.57 & 0-0 & R$_\mathrm{1f}24.5$ \\ 
820.4937 & 0.68: & 0.72: & 7.9:  
& 2.981 & -1.50 & 3-3 & R$_\mathrm{2f}29.5$ \\
821.1968  & 0.25: & 0.57: & 7.5:  
& 2.983 & -1.50 & 3-3 & R$_\mathrm{2e}29.5$ \\
\noalign{\smallskip}
  \hline
  \end{tabular}
  \end{center}
  \smallskip
\end{table*}

\subsection{Pure Rotation OH Lines}

Observed, mid-infrared, pure rotation lines are summarized in
Tables \ref{OH_3} and \ref{OH_2}. Laboratory line positions are
accurate to about $0.0004\,\mbox{cm$^{-1}$}$ or $0.15
\,\mbox{km\,s$^{-1}$}$ \citep{gold}. We use line data from the
\citet{gold} line list and the assignments in the Table are from
the mid-infrared Atlas of the sunspot spectrum
\citep{solspectrum}. The continuum
  around $919.7\,\mbox{cm$^{-1}$}$ is not well defined due to
  spurious features due to telluric absorption there, severely affecting the measurements.
  The observed lines are
at the photospheric velocity. For the three observational sets the
values are: $\langle\Delta v_\mathrm{OH(810)}\rangle$ = $-0.4 \pm
0.2\pm0.5\,\mbox{km\,s$^{-1}$}$, $\langle\Delta
v_\mathrm{OH(890)}\rangle$ = $1.4 \pm
0.1\pm1.5\,\mbox{km\,s$^{-1}$}$, and $\langle\Delta
v_\mathrm{OH(910)}\rangle$ = $-0.7  \pm
0.2\pm0.5\,\mbox{km\,s$^{-1}$}$, see Table \ref{OH_3} and
\ref{OH_2}. After the standard deviation of the mean, the
estimated systematic uncertainties are given. The latter is
generally estimated to $\pm0.5\,\mbox{km\,s$^{-1}$}$, but
the OH (and H$_2$O) lines from the February 4 spectra are shifted
by about $1.5 \,\mbox{km\,s$^{-1}$}$ relative to the photospheric
velocity, a shift not seen in the February 2 and 3 observations.
We, therefore, suppose that the anomalous shift arises from the
wavelength calibration of the February 4 spectra.

The measured line widths of the measured $1-1$, $2-2$, and $3-3$
OH lines give a mean FWHM of $8.8 \pm 0.3\,\mbox{km\,s$^{-1}$}$.
The instrumental profile corresponds to a FWHM of about
$3.7\,\mbox{km\,s$^{-1}$}$. Correcting for this, the intrinsic
width of these OH lines is about $7.9\pm
0.4\,\,\mbox{km\,s$^{-1}$}$, which is in fair agreement with the
width of the vibration-rotation OH lines.

The photosphere is certainly home to the OH molecules giving the
vibration-rotation lines. By velocity and FWHM similarities, the
suspicion is strong that the pure rotation lines  are also of
photospheric origin. Below (Sec. \ref{resluts}) we further the
argument for this origin by comparing synthetic and observed
spectra.



\subsection{H$_2$O Lines}

A call for adequate representation of the line opacity from H$_2$O
molecules needed  in construction of model photospheres of cool
stars has been answered by the construction of line lists using
techniques of theoretical molecular spectroscopy. For example, the
{\sc scan} list \citep{uffe_H2O} contains 100 million lines with
predicted wavelengths and line strengths. \citet{par} combine
predictions with laboratory measurements. Available line lists are
discussed by J\o rgensen et al. (2001) and \citet{bernath_rev}.
Present line lists are likely satisfactory for an opacity
calculation and probably for comparing synthetic and observed low
resolution spectra (see, for example, Jones et al. 2001 and Ryde
\& Eriksson 2002)\nocite{jones,ryde_eriksson}, but our need for
precise line positions cannot be met by theoretical calculations.
Partridge \& Schwenke's calculations match laboratory measurements
of line positions to about $0.05\,\mbox{cm$^{-1}$}$ or about $20
\,\mbox{km\,s$^{-1}$}$ for the lines of interest to us.

Therefore, we take line positions from laboratory measurements of
strong H$_2$O lines \citep{poly_1, poly_2, poly_3}. These
measurements made at a resolution of $\Delta\tilde\nu \approx
0.01\,\mbox{cm$^{-1}$}$, provide wavenumbers, excitation energies
of the lower state of the transition, and quantum state
assignments (with the rotational quantum numbers $J$, $K_a$, and
$K_c$\footnote{$K_a$ is the approximate projection of the total
angular momentum, $J$, along the rotational axis with the smallest
momentum of inertia and $K_c$ is the corresponding projection
along the rotational axis with the largest momentum of inertia.})
of the ground vibrational state, $(v_1 v_2 v_3)=(000)$, and the
first excited bending vibrational state, $(v_1 v_2 v_3)=(010)$.
These are states that give rise to relevant transitions for our
study. The uncertainty in the line positions is less than
$0.002\,\mbox{cm$^{-1}$}$
(P. Bernath 2002, private communication). By using the excitation
energies of the measured lines, we are able to identify the
corresponding lines in the line list by \citet{par}, which
provides us with accurate transition probabilities ($gf$ values).
The excitation energies (to be used in the calculation of the
level populations) in \citet{poly_1, poly_2} and \citet{par} agree
excellently with the experimentally measured energy levels of
\citet{water_E}. We use the partition function for water from
\citet{par} in our calculations of the synthetic spectra. Our new
list of water-vapor lines is presented in Tables \ref{water_12um}
and \ref{water_11um}. The line positions and assignments are given
as in Polyansky et al. (1996, 1997a). The $gf$ values and the
excitation potentials are taken from \citet{par}. In order to test
the wavelength accuracy of our new line list,
we compared with the sunspot spectrum \citep{solspectrum} finding
a good agreement within the uncertainties of the line lists.
We are therefore confident that the wavelengths of the few
water-vapor lines, used in the compilation of our new line list,
are accurate.

Single, unblended H$_2$O lines are expected to be the rule when
H$_2$O contributes weakly, as in the Arcturus spectrum.
But
there are certainly blends of water-vapor lines not accounted for
in our line list. Indeed, in the \citet{par} line list, we
detected a number of lines sufficiently strong to affect the
synthetic spectrum within one or a few resolution elements of a
measured line (assigned by Polyansky and collaborators).
We, therefore, also included these lines in our list with the
(uncertain) wavelength shifts found in the \citet{par} list.
In Tables \ref{water_12um} and \ref{water_11um} these extra lines
are given with only 6 digits, whereas the \citet{poly_1, poly_2}
lines are given with 8 digits (as is given in these references).
Some missing lines and blends may also  affect our synthetic
spectrum. We also note that the Partridge \& Schwenke line list
shows artificial splittings of a large number of degenerate,
rotational states, which could make assignments difficult
\citep[see the discussions in][]{poly_3, jones, ryde_eriksson}.


In Tables $\ref{water_12um}$ and $\ref{water_11um}$ measurements
of stellar water-vapor lines are shown.
The mean velocity shifts of the observed lines relative to the
stellar velocity are negligible: $\langle\Delta
v_\mathrm{(810)}\rangle$ = $0.1 \pm
0.2\pm0.5\,\mbox{km\,s$^{-1}$}$, $\langle\Delta
v_\mathrm{(890)}\rangle$ = $1.6 \pm
0.2\pm1.5\,\mbox{km\,s$^{-1}$}$, and $\langle\Delta
v_\mathrm{(910)}\rangle$ = $-0.1 \pm
1.1\pm0.5\,\mbox{km\,s$^{-1}$}$. In comparing the velocity shifts
of the OH and water lines in the $800-822\cm1$ region (which has
most measured values), we note that the systematic, instrumental
shifts will be the same for both species. Thus, there might be a
small red-shift of the H$_2$O lines relative to the OH lines of
$0.5\pm0.3\kms$. A velocity shift of this order is indeed expected
between lines formed at different depths in the photosphere in a
star such as Arcturus, see for example \cite{allende}.
The FWHM of the water-vapor lines is measured to be
$\langle\mathrm{FWHM}\rangle= 8.2 \pm 0.3\,\mbox{km\,s$^{-1}$}$.
Correcting for the intrinsic line width, we find
$\langle\mathrm{FWHM}\rangle= 7.3 \pm 0.4\,\mbox{km\,s$^{-1}$}$,
which is slightly lower than the line widths found for the OH
lines.
Thus, the water-vapor lines, by their line width and velocity,
appear to be formed in the photosphere

\section{MODEL PHOTOSPHERES AND THE GENERATION OF SYNTHETIC SPECTRA}

For the purpose of analyzing our observations, we have generated
synthetic spectra based on model photospheres calculated with the
latest version of the {\sc marcs} code. This version of the {\sc
marcs} code is the final major update of the code and its input
data in the suite of {\sc marcs} model-photosphere programs first
developed by \citet{marcs:75} and further improved in several
steps, e.g. by \citet{plez:92}, \citet{jorg:92}, and
\citet{BDP:93}.

The {\sc marcs} hydrostatic, spherical model photospheres are
computed on the assumptions of Local Thermodynamic Equilibrium
(LTE) including chemical equilibrium, homogeneity and the
conservation of the total flux (radiative plus convective; the
convective flux being computed using the mixing length
formulation). The radiative field used in the model generation, is
calculated with absorption from atoms and molecules by opacity
sampling at approximately 84\,000
wavelength points over the wavelength range $2300\,\mbox{\AA} $--$
20\,\mbox{$\mu$m}$.

 Data on the absorption by atomic species are collected from the
VALD database \citep{VALD} and Kurucz (1995, private
communication). The opacity of CO, CN, CH, OH, NH, TiO, VO, ZrO,
H$_2$O, FeH, CaH, C$_2$, MgH, SiH, and SiO are included and
up-to-date dissociation energies and partition functions are used.
The continuous absorption as well as the new models will be fully
described in a series of forthcoming papers in A\&A (Gustafsson et
al., J\o rgensen et al., and Plez et al., all in preparation).

The model photosphere is described, for every depth point through
the photosphere, by the radius ($R$), the standard optical depth
calculated at 500 nm ($\tau_{500}$), temperature ($T$), electron
pressure ($P_e=N_e \mathrm k T$), gas pressure ($P_g=N  \mathrm k
T$), density ($\rho$), and standard opacity at $500\,\mbox{nm}$.
Normally, the models are calculated with 54 depth points from
$\log \tau_\mathrm{Ross}=2.0$ out to $\log
\tau_\mathrm{Ross}=-5.6$, which in our case corresponds to an
optical depth evaluated at $5000\,\mbox{\AA }$ of $\log
\tau_{500}=-4.8$. The physical height above the $\log
\tau_{500}=0$ layer of this outermost point is
$6.2\times10^{10}\,\mbox{cm}$ or $3\%$ of the stellar radius. We
will use the $\log \tau_{500}$ scale as a depth scale in the
following discussion.

Using the model photosphere we calculate synthetic spectra by
solving the radiative transfer in a spherical geometry. We
calculate the radiative transfer for points in the spectrum
separated by  $\Delta \tilde{\nu} \sim 1\,\mbox{km\,s$^{-1}$}$
(corresponding to a resolution of $\tilde{\nu}/\Delta \tilde{\nu}
\sim 330\,000$) even though the final resolution is lower.




Arcturus is a well studied star. The effective temperature derived
by \citet{griffin} is, for example, $T_\mathrm {eff}= 4290\pm
30\,\mbox{K}$. \citet{peterson} deduce its stellar parameters by
synthesizing the optical spectrum using extensive line lists, and
find $T_\mathrm {eff}= 4300\pm 30\,\mbox{K}$, $\log g =
1.5\pm0.15\,\mbox{(cgs)}$, [Fe/H]$=-0.5\pm0.1$, and
$\xi_t=1.7\pm0.3\,\mbox{km\,s$^{-1}$}$. \citet{leen_aBoo} derive
the stellar parameters from studying the entire Infrared Space
Observatory (ISO) spectrum ($2.4-12\,\mbox{$\mu$m}$) of Arcturus.
They used the same {\sc marcs} code and the same input data as we
are using. Their parameters for Arcturus are $T_\mathrm {eff}=
4300\,\mbox{K}$, $\log g = 1.75\,\mbox{(cgs)}$,
$\xi_t=1.7\,\mbox{km\,s$^{-1}$}$, $M=1\,\mbox{M$_\odot$}$, and the
specific abundances $\epsilon(\mathrm C)=7.90$, $\epsilon(\mathrm
N)=7.55$, and $\epsilon(\mathrm O)=8.67$ on the usual logarithmic
scale of 12. The surface gravity of \citet{leen_aBoo} seems to be
somewhat high in comparison with other literature values. We will,
therefore, calculate our standard model with $T_\mathrm {eff}=
4300\,\mbox{K}$, $\log g = 1.5\,\mbox{(cgs)}$, [Fe/H]$=-0.5$, and
$\xi_t=1.7\,\mbox{km\,s$^{-1}$}$, and use the abundances of
carbon, nitrogen, and oxygen given by \citet{leen_aBoo}.

We find that the temperature structure differs between a model
atmosphere calculated in plane-parallel geometry and a model in
spherical geometry for $\tau_{500}\lesssim-1$, with a maximum
difference of $65\,\mbox{K}$ at $\tau_{500}\sim-4$. To minimize
the uncertainties we therefore choose to calculate our models in
spherical geometry.

In order to test the stellar model we start by synthesizing
Arcturus' spectrum at
$6400\,\mbox{cm$^{-1}$}$ and confront it with the high-resolution
observations presented in the infrared Arcturus Atlas
\citep{arcturusatlas,arcturusatlas_II}, see Figure \ref{fig4}.
The synthesized region, $6350-6430\,\mbox{cm$^{-1}$}$, is chosen
since it contains several first-overtone, vibration-rotation OH
lines ($v'-v''= 2-0$ and $3-1$).
The synthetic OH lines are narrower than their real counterparts.
To match the observed widths, we introduce the customary artifice
of a macroturbulent velocity. We suppose these velocities follow a
Gaussian distribution.  The adopted macroturbulence parameter is
$5.7\pm0.5\kms$. Figure \ref{fig4} shows that OH lines are fitted
well with our model, indicating the relevance of the model
photosphere and the stellar parameters used in modelling it.
No attempt has been made to fit the
metal lines, nor has an attempt been made to make a complete line
list for the region. The focus has been to show that the
vibration-rotation OH lines are ascribed to the star's
photosphere. This demonstration is a firm basis from which to
begin interpretation of the pure rotation OH and H$_2$O lines.













\begin{figure}
\caption{Pure rotation, OH($v=0-0$) lines. The
observed spectrum is plotted with crosses. The dashed line shows
the synthetic spectrum calculated based on our original model
photosphere. The full line shows the spectrum based on the
photosphere with the new temperature structure. \label{fig7}}
\end{figure}

\begin{figure}
\caption{Pure rotation, OH($v=1-1$) lines. See
Figure \ref{fig7} for an explanation for the different curves. The
line at $918.8\cm1$ contains a water-vapor blend. \label{fig8}}
\end{figure}

\begin{figure}
\caption{Pure rotation, OH($v=2-2$) lines.  See
Figure \ref{fig7} for an explanation for the different curves.
\label{fig9}}
\end{figure}




\begin{figure}
\caption{Pure rotation,
water-vapor lines. The observed spectrum is plotted with crosses.
The dashed line shows the synthetic spectrum calculated based on
our original model photosphere. The full line shows the spectrum
based on the photosphere with the new temperature structure.
\label{fig11}}
\end{figure}




Confident of the stellar parameters used in the generation of the
model photosphere, we now confront the mid-infrared,
high-resolution {\sc texes} spectra with a synthetic spectrum of
the
$806-923\,\mbox{cm$^{-1}$}$ region, containing pure rotation OH
lines and H$_2$O lines.  Observed and model spectra are shown in
Figures $\ref{fig7}, \ref{fig8}, \ref{fig9},$ and $\ref{fig11}$.
The dashed line shows the synthetic spectra calculated with our
model-photosphere.
The wavenumber scale is expressed in the stellar rest frame. The
synthetic, near-infrared spectrum of Arcturus was convolved with
Gaussians, representing macroturbulence and the instrumental
profile of $6\kms$. For the mid-infrared spectra, a
macroturbulence parameter of $4.7\pm0.5\kms$ fits the OH lines
well.  To within the uncertainties, these estimates from different
spectral regions and spectra obtained at different times are
identical.

Given that the model and adopted abundances give an excellent fit
to the 1.6 $\mu$m vibration-rotation OH lines, it is of interest
to see if the same model-abundance combination fits the pure
rotation OH lines. The answer (Figures $\ref{fig7}-\ref{fig9}$) is
that the OH($1-1$), OH($2-2$), and OH($3-3$) lines in the
synthetic spectra are a very good fit to the observed lines. The
quality of the fit is poorer for the OH($0-0$) lines in that the
predicted lines are weaker than observed.

In comparison to the mild failure to predict the $0-0$ OH lines,
the failure to account for the observed H$_2$O lines is dramatic
(Figure $\ref{fig11}$).  It is impossible to find a {\sc marcs}
model that would predict the line strengths of all H$_2$O and OH
lines by only changing the effective temperature and surface
gravity. Some water-vapor lines can by fitted by a
$4100\,\mbox{K}$ model whereas others can only be synthesized with
a $3900\,\mbox{K}$ model. The OH($v=0-0$) lines need even lower
effective temperatures: approximately $3000\,\mbox{K}$. Higher
surface gravities tend to strengthen the lines. Such a failure
must hold novel information about Arcturus' atmosphere. Given that
the H$_2$O lines are at the photospheric velocity and have the
width of the photospheric OH vibration-rotation and pure rotation
lines, we argue that the H$_2$O lines are also photospheric lines.
It should also be noted that circumstellar emission or infrared
excess of thermal radiation from dust is not seen from Arcturus, a
conclusion demonstrated by the ISO fits of the star's
$2.4-12\,\mbox{$\mu$m}$ spectrum \citep{leen_aBoo}.
Thus, the failure is taken to be a signal that the model
photosphere and/or the synthetic spectrum is an inadequate
representation of the real photosphere. Were the H$_2$O lines
formed in a detached layer such as a MOLsphere it seems likely
that their velocity, line width, and degree of excitation would
not match so closely the photospheric values.








\begin{figure}
\caption{The temperature and gas-pressure
structures of our models. The full line shows the original {\sc
marcs} model. The dotted line shows the same model but from a new
calculation extending it further out. The dashed line shows our
model that fit our observations, with a modified T-structure at
$\log \tau_{\mathrm {500}}<-3.8$ and beyond. The electron and gas
pressures and the density are calculated consistently from this
temperature structure. The apparent difference in the gas-pressure
structure between the original model (full line) and its extension
(dotted line) is most likely a numerical boundary effect in the
original model.\label{fig15}}
\end{figure}

\section{DISCUSSION\label{resluts}}

The presence of a detectable column density of H$_2$O in Arcturus'
photosphere is at odds with our predictions which combine an
appropriate {\sc marcs} model with a spectrum synthesis code. One
or more aspects of the combination must be in need of revision. In
this section, we comment on some possible revisions.

\subsection{A Cooler Upper Photosphere?}

Perhaps, the simplest revision providing stronger predicted H$_2$O
lines is to drop the temperatures in the uppermost, uncertain
layers of the model photosphere.
{\sc marcs} models are normally begun at $\log\tau_{500} = -4.8$
and extend into the star. To assure ourselves that an outward
extension would not affect the synthetic spectra, we recomputed
the {\sc marcs} model with the first depth point at
$\log\tau_{500}$ = $-$7.0, corresponding to
$8.8\times10^{10}\,\mbox{cm}$, or $4\%$ of the stellar radius,
above the $\log \tau_{500}=0$ layer. The OH and H$_2$O lines in
the synthetic spectrum are almost unchanged; the predicted H$_2$O
lines remain far weaker than observed. A helpful reference mark in
discussions of model structure is the location of optical depth
unity in the mid-infrared continuum: $\tau_{12 \rm{\mu m}} = 1$
occurs at $\log\tau_{500} = -$0.5.

After one or two tests, it became clear that a minor revision of
temperatures in the extreme boundary layers suffices to impose
strong H$_2$O lines on the synthetic spectrum. By cooling the
temperature structure, water vapor starts to form in the outer
layers. We find that the strengths of the water lines are very
sensitive to the temperature structure of the very outer
photosphere.
In Figure \ref{fig15}, we show an alternative choice for the
temperature in this boundary layer with temperatures cooler than
the {\sc marcs} model for $\log\tau_{500} < -3.8$; the model now
violates the condition of flux constancy imposed on {\sc marcs}
models. Over most of the region, the temperatures are a mere
$300\,\mbox{K}$ cooler than the {\sc marcs} model. Electron and
gas pressures were recalculated assuming hydrostatic equilibrium.
The density and standard opacity are also consistently
recalculated from this new temperature structure, as are the
ionization balance, molecular equilibria,
and opacities. New synthetic spectra were calculated assuming, as
before, LTE for the molecular equilibrium and the line formation.

The new spectra (solid lines in Figures $\ref{fig7}-\ref{fig11}$)
provide a pleasing simulation of the observed  OH and H$_2$O
lines. The OH vibration-rotation lines at $6400\cm1$ and the OH
pure rotation $3-3$, $2-2$, and $1-1$ lines are unaffected by the
boundary layer revisions. The OH pure rotation $0-0$ and H$_2$O
lines are strengthened over the predictions from the {\sc marcs}
model in such a way that the observed lines are well fit.


A lowering of the boundary layer temperatures of Arcturus was
proposed by \citet{ke3} who analyzed high-resolution spectra of CO
fundamental vibration-rotation lines at $4.6\,\mbox{$\mu$m}$. To
explain the line depths of the strongest CO lines, Wiedemann et
al. discarded a chromospheric temperature rise (which starts off
at $\tau_{500}\sim -3$)
and required the boundary temperature to fall to $2400\,\mbox{K}$
in the highest layers contributing the CO line cores. Their
synthetic spectra were computed on the assumption of non-LTE for
the excitation of the CO molecule. The temperatures in the
boundary layer needed to fit the CO line profiles  are dependent
on the choice of (poorly known) collision cross-sections needed in
the non-LTE calculation. For this and other reasons (e.g., we
assume the OH and H$_2$O molecules are in LTE), quantitative
agreement between the temperature profile in Figure \ref{fig15}
and the CO-based result is not expected. The boundary temperatures
we derive are a few hundred degrees lower, but our model predicts
the CO lines at $4.6\,\mbox{$\mu$m}$ just as well.

The revisions to the upper boundary layer made by us
were made solely with the aim of fitting observed line
profiles. Here, we comment briefly on possible physical processes
behind a cooling of the boundary layer.

How accurately are the boundary layer temperatures predicted by
the {\sc marcs} code? Assuming the underlying assumptions are
valid, we may break the sources of uncertainty into two classes:
those related to the defining atmospheric parameters, and those
associated with the atomic and molecular data, principally the
line blanketing. The atmospheric parameters of Arcturus are quite
well determined based on a range of different observing
techniques. No obvious source of line blanketing is discarded from
the model calculation. However, there may be uncertainties in the
input data.





The numerical uncertainties are certainly less than
$50\,\mbox{K}$. One aspect of the statistical uncertainties in the
modelled temperature structure may be estimated by selecting
different numbers of points of the opacity sampling (OS) in the
model calculation. The outer-most layers of the model will be
sampled only by CO bands and the strongest metal lines. The
temperature difference between using $2,100$ and $11,000$ OS
points is at most 30 K and in the outer layers at most 10 K.
Similarly, between using $2,100$ and the full $84,000$ OS points,
the maximum difference in the temperature structure is 40 K, and
in the most shallow parts also 10 K.

Relaxation of the  assumptions behind the {\sc marcs} models may
also change the atmospheric structure. Even small alterations of
the heating or cooling terms in the energy equation (for example
due to dynamic processes or uncertainties and errors in the
calculations of radiative cooling) may lead to changes in the
temperature structure in the outer, tenuous regions of the
photosphere where the heat capacity per volume is low. The
assumption of LTE is one target of model builders seeking to go
beyond the standard assumptions. Atomic -- continuous and line --
opacity was treated in non-LTE using large model atoms by Short et
al. (2002, in preparation) in spherical geometry, hydrostatic
equilibrium, and flux constancy. Introduction of non-LTE in place
of LTE lowered the boundary temperature by an amount quite similar
to our empirical cooling of the {\sc marcs} model. This suggests
that non-LTE effects in atoms may suffice to account for the
surprising presence of H$_2$O molecules in the photosphere.



\subsection{An Inhomogeneous Atmosphere?}

A key assumption made in construction and application of the model
atmosphere is that physical conditions are everywhere the same at
a given geometrical depth. Such a model is said to be
`homogeneous'. Arcturus, with its deep convective envelope, is
certain to be inhomogeneous, that is to exhibit surface
granulation \citep{svart} which may be crudely characterized as
consisting of columns of warm rising and cool sinking gas. The
continuum intensity contrast in the mid-infrared will be small.
There will, however, most likely be a sharp difference in the
H$_2$O column density with the sinking column much richer in the
molecules, and, as a result, the surface-averaged spectrum will
show stronger H$_2$O lines than the spectrum of the equivalent
homogeneous model atmosphere. Unfortunately,  inhomogeneous model
atmospheres of giant stars are unavailable, but we note that
progress is being made \citep{bf}. Yet, the modest 300 K cooling
imposed on the {\sc marcs} model is surely within the range of
temperature differences between rising and sinking granules. Note
that granulation is likely to have a weaker effect on the CO lines
because carbon atoms are quite fully associated into CO molecules.
Contrast between the CO column density for the rising and sinking
granules will be less than for H$_2$O.



An alternative mechanism for producing inhomogeneities is
variously known as `a molecular catastrophe' or `temperature
bifurcation'.
 In warm stars, the CO
molecular opacity influences the temperature profile predicted by
a program like {\sc marcs}.
 If the temperature of gas in the boundary layer
is perturbed, there may be a runaway to one of two different
solutions. Suppose the gas is cooled, more CO molecules form and
increase the cooling of
 the
boundary layer. The temperature drop leads to yet more CO
molecules and to a runaway to a `cool' boundary layer. Conversely,
if the perturbation raises the temperature, CO molecules are
dissociated, the cooling rate  decreased, and the temperature rise
is continued.  In cool O-rich stars, where the CO number density
is limited by the abundance of carbon, this sensitivity of opacity
to temperature is greatly reduced, but other molecules (e.g., SiO
or H$_2$O) may act in a similar way.
 Quite detailed studies of the bifurcation induced by CO molecules have
been made for the Sun and similar stars
\citep{muchmore_ulm,muchmore,muk}. Supporting evidence has come
from observations of CO $4.6\,\mbox{$\mu$m}$ lines at high
spectral resolution \citep{at,atb}. Convective velocity flows are
likely to hamper the development of the temperature bifurcations
\citep{steffen_much}.

Arcturus certainly exhibits a chromosphere, which has been
modelled by \citet{al:75}. Limits on its filling factor are
discussed by \citet{ke3} based on their deduced temperature
structure. It is suggested that the hot component can not be
larger than few percent. Existence of the chromosphere around
Arcturus suggests a heating mechanism active in and above the
upper photosphere. \citet{cuntz_1} made a preliminary study of the
effects of propagating acoustic waves including molecular line
cooling. When the waves lead to weak shocks, a cool atmosphere is
formed: temperatures as low as 1000 K were predicted. These
temperatures are substantially below those required to account for
the H$_2$O lines and the CO $4.6\,\mbox{$\mu$m}$ lines
\citep{ke3}. On the other hand, strong shocks lead to a
temperature inversion, approximately matching an empirical
chromospheric model for Arcturus. (The amplitude and period of the
investigated waves are a poor match to the observed radial
velocity variations \citep{hc}.) These calculations incorporating
a driving mechanism for temperature bifurcation support the idea
of an inhomogeneous upper photosphere and low chromosphere.



\subsection{Non-LTE  Effects}

Order of magnitude estimates of collisional and radiative rates
indicate that the vibronic structure of the OH and H$_2$O
molecules are unlikely to be in LTE throughout the boundary layer.
To illustrate this claim, we compare rates of de-excitation
through collisions and photons.

Collisional de-excitation occurs at a rate $C_\mathrm{ul} \sim
N_\mathrm H\sigma_\mathrm{ul} v_\mathrm{rel}$ where we consider H
atoms at density $N_\mathrm H$ to be the dominant collision
partner, the cross-section $\sigma_\mathrm{ul}$ is taken as the
geometrical area ($\sim 10^{-15}$ cm$^{-2}$), and $v_\mathrm{rel}$
is the relative velocity between the H atom and the molecule
($\sim 5$ km s$^{-1}$). To maintain LTE a minimum condition is
that the rate $C_\mathrm{ul}$ must exceed the rate of spontaneous
emission given by sum of the Einstein coefficients
($A_\mathrm{ul}$) from the upper level. For the OH molecule, the
$A$-values of rotational levels of the lowest vibrational level
($v^\prime$ = 0) run from about 2 s$^{-1}$ at low $J$ to about
1000 s$^{-1}$ at $J \simeq 40$ scaling approximately as $J^2$
(Goldman et al. 1998). For other vibrational levels, the  values
are of a similar magnitude even when the contribution of
vibration-rotation transitions are included in the deexcitation.
A-values of the H$_2$O levels, for the lines of interest, range
from $10$ to $1000\,\mbox{s$^{-1}$}$.

Supposing the sum of the $A$ values to be $A_\mathrm{sum} =$
 1, 100, and 1000 s$^{-1}$, $C_\mathrm{ul} \sim
A_\mathrm{sum}$ for gas pressures $\log P_g \sim -3.2$, $ -1.2$,
and $-0.2$ (cgs) which according to Fig. \ref{fig15} correspond,
for the first value, to an optical depth beyond our boundary
point, and for the second and third ones, to $\log\tau_{500} \sim
-6$ and $-5$, respectively. For example, for the water-vapor line
at $818.4\,\mbox{cm$^{-1}$}$, with an $A_\mathrm{ul}$ of
$600\,\mbox{s$^{-1}$}$, we find a critical
gas pressure
of $\log P_g\sim -0.4$ (cgs) which occurs already at $\log
\tau_{500}\sim -5.3$.
Hence, we cannot be confident that collisions will dominate over
radiative decay and therefore we can not conclude, on the basis of
this calculation, that LTE necessarily prevails for the line
formation in the very outer layers of the photosphere.
The effect on line depths is difficult to determine. A proper
non-LTE calculation is necessary to determine whether LTE does in
fact prevail and what the effects would be due to a departure from
LTE.
To complete the assessment of non-LTE effects, one should study
the chemistry, that is the radiative and collisional processes
controlling the density of OH and H$_2$O molecules. In general,
these rates are expected to be lower than those controlling the
excitation.

\section{CONCLUDING REMARKS}



We have made an unexpected discovery of absorption lines of water
vapor in the disk-averaged spectrum of Arcturus. Based on
kinematic information, line widths and excitation temperature, we
argue that the water vapor is photospheric and not circumstellar.
Until now, H$_2$O has not been expected to exist in photospheres
of stars of Arcturus' effective temperature.
This is true unless the outer layers of the model photosphere do
not describe these regions properly. By modifying these layers we
have succeeded in finding {\it a} model which yields a synthetic
spectrum matching the observations. We are, however, not claiming
that this necessarily is the only and true model. Our exercise
simply shows that it is possible to achieve a photospheric
spectrum containing water vapor also for early K giants if the
outer parts are cooled for some reason. Possible reasons for the
unexpected water vapor include photospheric inhomogeneities (such
as convective flows, molecular catastrophes, or even star spots)
and departures from LTE in the photospheric structure or line
formation in the boundary layers.

We note that it is obvious that the chromospheric model proposed
for Arcturus \citep{al:75} would not fit our mid-infrared
observations. This circumstance was also found by \citet{ke3}
analyzing near-infrared CO data. Wiedemann  et al. conclude that
the upper photosphere has to be inhomogeneous with a quiet and a
chromospheric part.
Perhaps, the chief challenge reinforced by our detection of water
vapor concerns development of tools with which to construct
empirical atmospheres of an inhomogeneous atmosphere. The
mid-infrared H$_2$O and OH lines, also the cores of the 4.6 $\mu$m
vibration-rotation lines, are useful probes of the cooler gas in
the upper photospheric layers. Ultraviolet emission lines may be
the best probe of the hotter gas in these (and higher) layers.
What happens at several scale heights above these layers? One
possibility is that  deposition of mechanical energy heats the
cool upflowing gas and largely erases the distinction between
`hot' and `cool' gas. With the drop in density and the concomitant
increase in cooling times, the gas may remain warm to very large
distances from the red giant.





\acknowledgments

We should like to thank Drs. Kjell Eriksson and Bengt Gustafsson
for fruitful discussions, Dr. Peter F. Bernath for providing us
with his recent review on the spectroscopy of water vapor, Dr. Ian
Short for providing his Arcturus model, and Dr. Charles H. Townes
for piquing our interest in observations of water vapor in cool
stars. This work was supported by the Swedish Foundation for
International Cooperation in Research and Higher Education,
Stiftelsen Blanceflor Boncompagni-Ludovisi, n\'ee Bildt, the Royal
Physiographic Society of Lund, Sweden, and the Robert A. Welch
Foundation of Houston, Texas. The construction of the {\sc texes}
spectrograph was supported by grants from the {\sc nsf}, and
observing with {\sc texes} was supported by the Texas Advanced
Research Program. We acknowledge the support of Thomas K.
Greathouse and by the {\sc irtf} staff.




\bibliographystyle{aabib99}


\end{document}